\documentstyle[aabib,psfig]{l-aa} 

\newcommand{\msun}{\mbox{${\rm M}_\odot$}}

\def\apgt{\ {\raise-.5ex\hbox{$\buildrel>\over\sim$}}\ }
\def\aplt{\ {\raise-.5ex\hbox{$\buildrel<\over\sim$}}\ }

\begin{document}

%%\baselineskip 1.5 \baselineskip

\thesaurus{02.01.2;
	   02.02.1;
	   08.02.1;
	   08.14.1;
	   13.07.1;
	   13.07.3}

\title{Explaining the light curves of Gamma-ray
       Bursts with a precessing jet}

\author{Simon F.\ Portegies Zwart\thanks{Hubble Fellow}}
%	Chang-Hwan Lee\altaffilmark$^2$ 
%	and
%	 Kyu Lee\altaffilmark$^{2, 3}$
%}
\offprints{Simon Portegies Zwart: spz@grape.c.u-tokyo.ac.jp}
\institute{Department of Information Science and Graphics, 
		College of Arts and Science, 
		University of Tokyo, 3-8-1 Komaba,
		Meguro-ku, Tokyo 153, Japan}

%$^1$ Department of Information Science and Graphics, 
%		College of Arts and Science, 
%		University of Tokyo, 3-8-1 Komaba,
%		Meguro-ku, Tokyo 153, Japan
%
%$^2$ Department of Physics and Astronomy,
%		State University of New York at Stony Brook,
%                Stony Brook, NY11794, USA 
%
%$^3$ Department of Physics, Hanyang University, Seoul 133-791, Korea

\date{Received; Accepted:}
\maketitle
\markboth{Portegies Zwart, S.\ F. : Gamma-ray Binaries}{}

\maketitle

\begin{abstract}
A phenomenological model is presented to explain the light curves of
gamma-ray bursts. The model is based on a black hole which is orbited
by a precessing accretion disc which is fed by a neutron
star. Gamma-rays are produced in a highly collimated beam via the
Blandford-Znajek mechanism.  The beam sweeps through space due to the
precession of the slaved accretion disc. The light curves expected
from such a precessing luminosity cone can explain the complex
temporal behavior of bright gamma-ray bursts.

\end{abstract}

%%\vspace*{-3.5cm}

\section{Introduction}

Gamma-ray bursts are observed with a large variety in duration,
ranging from seconds to minutes (Norris et al.\
1996),\nocite{1996ApJ...459..393N} intensity and variability.  The
shortest temporal structures are unresolved by detectors and reflect
the activity of a highly variable inner engine (Fenimore et al.\ 1996).
On the other hand some bursts last for several minutes which indicates that
the energy generation within the burst region has a rather long time
scale.

\begin{figure*}
\centerline{\psfig{file=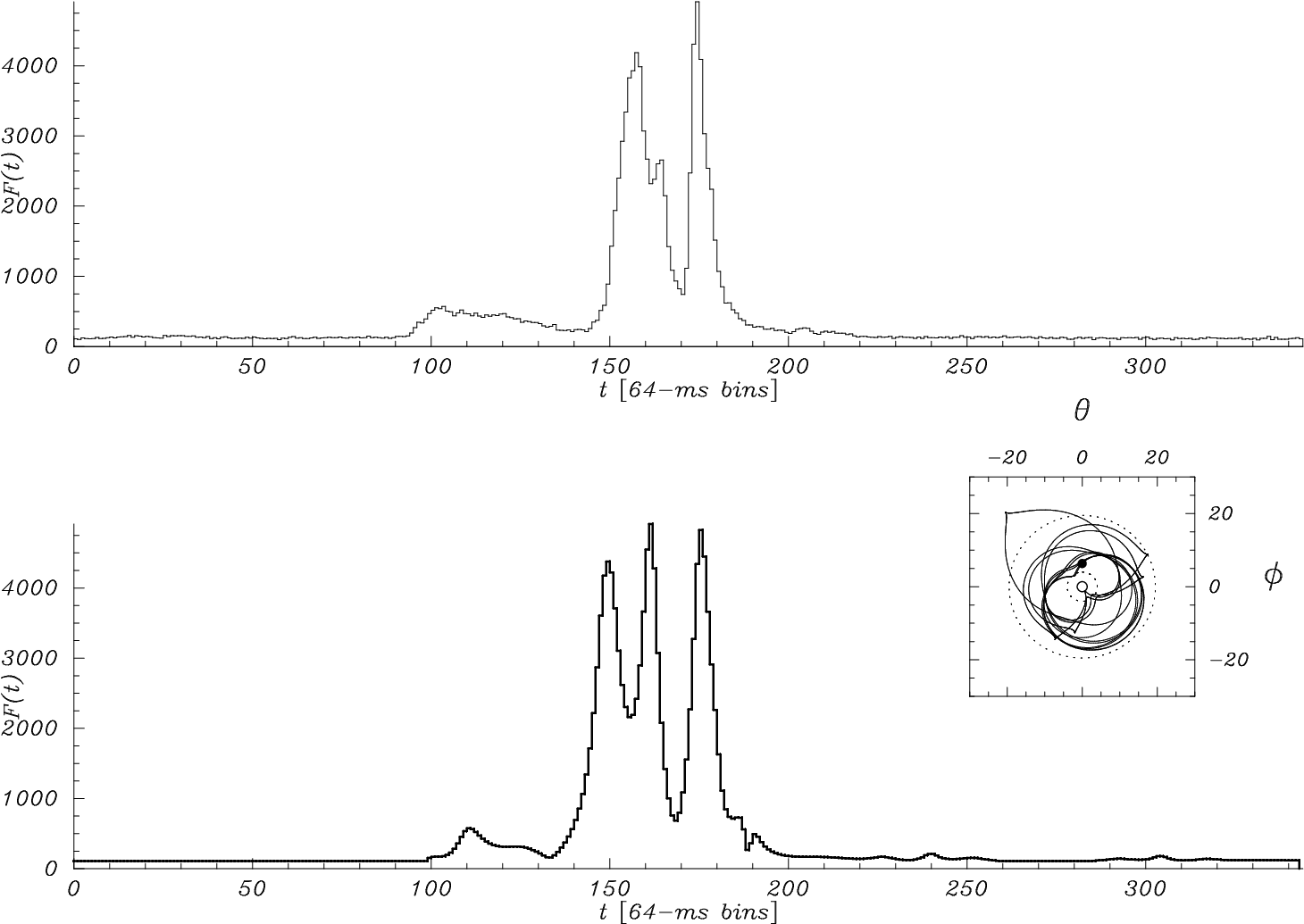,height=7cm,angle=-90}}
\caption[]{
The upper panel gives the flux (Y-axis) as a function of time (X-axis)
for  the gamma-ray burst with BATSE trigger numbers 1609. \\
The lower panel gives the result of the fitted burst. Time is in units
of 64\,ms, the time resolution of BATSE.  The 
upper right corner gives a schematic representation of the central
locus of the black hole (central $\circ$) and the trajectory of our
line of sight (solid line) starting at the $\bullet$, moving
clockwise. The inner dotted line identifies the angle at which the
luminosity distribution within the luminosity cone is maximum, it
drops to zero at the outer dotted line (see PZLL98) The simulated
burst is binned in 64\,ms bins to make a comparison with the
observation more easy.  }
\label{fig:GRB1609E3}
\end{figure*}

In the proposed model a neutron star transfers mass to a black hole
with a mass of 2.2 to 5.5\,\msun.  A strong magnetic field is anchored
in the disc, threads the black hole and taps its rotation energy via
the Blandford-Znajek (1977)\nocite{bz77} mechanism. Gamma-rays are
emitted in a narrow beam.  The luminosity distribution within the beam
is given by the details of the Blandford-Znajek process. Precession of
the inner part of the accretion disc causes the bean to sweep through
space. This results in repeated pulses or flashes for an observer at a
distant planet.

This model was proposed by Portegies Zwart et al.\ (1999, hereafter
PZLL) to explain the complex temporal structure of gamma-ray bursts.

\section{The Gamma-ray binary}

We start with a close binary where a low-mass black hole is
accompanied by the helium star. The configuration results from the
spiral-in of a compact object in a giant, the progenitor of the helium
star (see Portegies Zwart 1998, hereafter PZ).

The collapse of the helium core results in the formation of a neutron
star.  The sudden mass loss in the supernova and the velocity kick
imparted to the neutron star may dissociate the binary. If the system
remains bound a neutron star -- black hole binary is formed.

The separation between the two compact stars decreases due to
gravitational wave radiation (see \cite{PM63}).  At an orbital
separation of a few tens of kilometers the neutron star fills its
Roche-lobe and starts to transfer mass to the black hole.  Mass
transfer from the neutron star to the black hole is driven by the
emission of gravitational waves but can be stabilized by the
redistribution of mass in the binary system. If the mass of the black
hole $\aplt 2.2$\,\msun\, coalescence follows within a few orbital
revolutions owing the Darwin-Riemann instability (\cite{CE77}). If the
mass of the black hole $\apgt 5.5$\,\msun\ the binary is
gravitationally unstable (\cite{LS76}); the event horizon of the hole
is then larger than than the orbital separation. Only in the small
mass ranger from $2.2$\,\msun\ to $5.5$\,\msun\, stable mass transfer
is possible (see PZ).

The entire episode of mass transfer lasts for several seconds up to
minutes. Mass transfer becomes unstable if the neutron star starts to
expand rapidly as soon as its mass drops below the stability limit of
$\sim 0.1$\,\msun.  Initially the neutron star's material falls in the
black hole almost radially but at a later stage an accretion disc can
be formed (see PZ for details).

\section{The precessing jet}

The asymmetry in the supernova, in which the neutron star is formed,
causes the angular momentum axis of the binary to make an angle $\nu$
with the spin axis of the black hole. This misalignment causes the
accretion disc around the black hole to precess (see Larwood
1998).\nocite{Larw98}  

The magnetic field of the black hole is anchored in the disc.  Energy
can then be extracted from the black hole by slowing down its rotation
via the magnetic field which exert torque on the induced current on
the black hole horizon (Blandford \& Znajek 1977; Thorne et al.\
1986).\nocite{bz77}\nocite{tpm86} The radiation is liberated in a
narrow beam with an opening angle of $\aplt 6^\circ$\,(Fendt
1997).\nocite{fen97} Such a narrow beaming is supported by the results
of the ROTSE (Robotical Optical Transient Search Experiment) telescope
(see http://www.umich.edu/~rotse/ for more details).  Since the
magnetic field is anchored in the disc the radiation cone precesses
with the same amplitude and period as the accretion disc.

The intrinsic time variation of a single gamma-ray burst has a short
rise time followed by a linear decay (Fenimore
1997).\nocite{astro-ph/9712331} We construct the burst time profile
from three components: an exponential rise, a plateau phase and a
stiff decay (see PZLL for details). Rapid variations within this
timespan are caused by the
precession and nutation of the radiation beam. This results in a model
with eight free parameters; three timescales for the profile of the
burst, the precession and nutation periods, the precession angle and
the direction (two angles) from which the observer looks at the burst.

Figure\,\ref{fig:GRB1609E3} gives the result of fitting the model to
the gamma-ray burst with BATSE trigger number 1609 in the energy
channel between 115\,keV and 320\,keV. The fit was performed by
minimizing the $\chi^2$ with simulated annealing (see PZLL).  The
light curves computed with this model shows similar complexities and
variability as observed gamma-ray bursts (see
Fig.\,\ref{fig:GRB1609E3}). 

\acknowledgements 
I am grateful to Gerry Brown, Chang-Hwan Lee, Hyun Kyu Lee and Jun
Makino for a great time, numerous discussions and financial support. 

%\bibliographystyle{/usr2/spz/tex/lib/inputs/aabib}
%\bibliography{/usr2/spz/tex/lib/inputs/references,/usr2/spz/tex/lib/inputs/spz}

\begin{thebibliography}{9999}

%\bibitem[\protect\astroncite{Band \& Grindlay}{1984}]{bg84}
%Band, D.~L., Grindlay, J.~E. 1984, ApJ, 285, 702

%\bibitem[\protect\astroncite{Bagot et al.}{1998}]{BPZY98}
%Bagot, P., Portegies Zwart, S.\ F., \& Yungelson, L.\ R.\ A\&A 332, L57 

\bibitem[\protect\astroncite{Blandford \& Znajek}{1977}]{bz77}
Blandford, R.~D., Znajek, R. 1977, MNRAS, 179, 433

\bibitem[Clark et al.\ 1977]{CE77} Clark, J.\ P.\ A., \& Eardley, D.\
M.\ 1977, ApJ 215, 311 

%\bibitem[\protect\astroncite{{Falcke} \&
%  {Biermann}}{1998}]{1998astro.ph.10226F}{Falcke}, H., {Biermann}, P.~L. 1998,
%\newblock in A\&A

\bibitem[\protect\astroncite{Fendt}{1997}]{fen97}
Fendt, C. 1997, A\&A, 319, 1025

\bibitem[\protect\astroncite{{Fenimore}}{1997}]{astro-ph/9712331}
{Fenimore}, E.~E. 1997, astro-ph/9712331

\bibitem[\protect\astroncite{{Fenimore} et~al.}{1996}]{1996ApJ...473..998F}
{Fenimore}, E.~E., {Madras}, C.~D., {Nayakshin}, S. 1996, ApJ, 473, 998

%\bibitem[\protect\astroncite{{Hut} \& {Van Den
%  Heuvel}}{1981}]{1981A&A....94..327H}
%{Hut}, P., {Van Den Heuvel}, E. P.~J. 1981, A\&A, 94, 327

%\bibitem[Jaranowski \& Krolak 1992]{JK92} Jaranowski, P., \& Krolak,
%A.\ 1992, ApJ 394, 586 

%\bibitem[\protect\astroncite{Kobayashi et~al.}{1997}]{kps97}
%Kobayashi, S., Piran, T., Sari, R. 1997, ApJ, 490, 92

%\bibitem[Lai et al.\ 1993]{LRS93} Lai, D., Rasio, F.\ A., Shapiro, S.\
%1993, ApjS 88, 205

\bibitem[\protect\astroncite{{Larwood}}{1998}]{Larw98}
{Larwood}, J.~D. 1998, MNRAS, 299, L32

\bibitem[Lattimer \& Schram 1976]{LS76} Lattimer, J.\ M., \& , Schram 
D.\ M.\ 1976, ApJ 210, 549 

%\bibitem[\protect\astroncite{{Mao} \& {Paczynski}}{1992}]{1992ApJ...388L..45M}
%{Mao}, S., {Paczynski}, B. 1992, ApJ, 388, L45

%\bibitem[\protect\astroncite{{Norris} et~al.}{1996}]{1996ApJ...459..393N}
%{Norris}, J.~P., {Nemiroff}, R.~J., {Bonnell}, J.~T., {Scargle}, J.~D.,
%  {Kouveliotou}, C., {Paciesas}, W.~S., {Meegan}, C.~A., {Fishman}, G.~J. 1996,
%  ApJ, 459, 393
\bibitem[\protect\astroncite{{Norris} et~al.}{1996}]{1996ApJ...459..393N}
{Norris}, J.~P., {Nemiroff}, R.~J., {Bonnell}, J.~T., et al.\ 1996,
ApJ, 459, 393 

\bibitem[\protect\astroncite{Peters \& Mathews}{1963}]{PM63}
Peters, P.~C., Mathews, J. 1963, Phys. Rev. D, 131, 345

\bibitem[\protect\astroncite{Portegies Zwart}{1998}]{pz98}
Portegies Zwart, S.\ F. 1998, ApJ 503, L53, PZ 

%\bibitem[\protect\astroncite{Portegies Zwart \& Yungelson}{1998}]{pzy98}
%Portegies Zwart, S.\ F., Yungelson L.\ R. 1998, A\&A 332, 173 

\bibitem[\protect\astroncite{Portegies Zwart et al.}{1998}]{pzll98}
Portegies Zwart, S.\ F. Lee, C.-H., \& Lee, H.\ K.\ 1999, ApJ 520
(astro-ph/9808191), PZLL

%\bibitem[\protect\astroncite{Press et~al.}{1992}]{NumResp}
%Press, W., Teukolsky, S., Vettering, W.T., Flannery B.P.\ 1992,
%Numerical Recipes, Cambridge Univ. Press., Cambridge

%\bibitem[\protect\astroncite{Sari \& Piran}{1997}]{sp97}
%Sari, R., Piran, T. 1997, ApJ, 485, 270

\bibitem[\protect\astroncite{Thorne et~al.}{1986}]{tpm86}
Thorne, K., Prince, R., MacDonald, D. 1986,
Black Holes; The membrane paradigm,
Yale Univ. Press

%\bibitem[\protect\astroncite{{Van Den Heuvel} et~al.}{1980}]{1980A&A....81L...7V}
%{Van Den Heuvel}, E. P.~J., {Ostriker}, J.~P., {Petterson}, J.~A. 1980, A\&A,
%  81, L7

\end{thebibliography}

\end{document}